\documentclass[a4paper,twoside,11pt]{article}
\usepackage[it, bf]{caption}
\usepackage{graphics}
\usepackage{graphicx}
\usepackage{epsfig}
\usepackage{amsmath}
 \usepackage{txfonts}
\usepackage{titleps}
\newpagestyle{mystyle}{
  \setfoot[\thepage][][]{}{}{\thepage}
}
\begin{document}
\setcounter{page}{142}
\begin{center}
\Large{\textbf{A Broadband 3-D Numerical FEM Study on the Characterization of
Dielectric Relaxation Processes in Soils\footnote{Proc. 10th International
Conference on Electromagnetic Wave Interaction with Water and Moist Substances,
ISEMA 2013, Editor: K. Kupfer,
Weimar, Germany, Sep 25-27, 142-151, 2013}}}\vspace{0.5cm}\\
 \large{\textmd{Norman Wagner$^1$, Markus Loewer$^2$}}\vspace{0.2cm}\\
 $^1$\textit{ \large{{\textmd{Institute of Material Research and Testing at the Bauhaus-University Weimar, Germany}}}}\\
 $^2$\textit{ \large{{\textmd{Leibniz Institute for Applied Geophysics (LIAG), Hannover, Germany}}}}
 \vspace{0.05cm}
 \end{center}
\subsection*{\textit{ABSTRACT}}

Soil as a complex multi-phase porous
material typical exhibits several distributed relaxation processes in the frequency range from 1~MHz to
approximately 10~GHz of interest in applications. To relate physico-chemical
material parameters to the dielectric relaxation behavior, measured dielectric
relaxation spectra have to be parameterized. In this context, a broadband numerical 3D FEM
study was carried out to analyze the possibilities and limitations in
the characterization of the relaxation processes in complex systems.\\

\noindent{\it Keywords: soil matric potential, soil moisture, dielectric
relaxation behavior of soil}

\section*{\large{1 INTRODUCTION}}

The dielectric relaxation behavior of porous media such as soils contains
valuable information of the material due to a strong correlation with the
volume fractions of the soil phases as well as contributions by interactions
between the pore solution and mineral particles \cite{Wagner2011,
WagScheu2009a}. Hence, high frequency electromagnetic remote sensing techniques
offer the possibility to estimate physico-chemical parameters fast and
non-invasive \cite{Robinson2008}.

To relate physico-chemical material parameters to the dielectric
relaxation behavior, measured dielectric spectra have to be parameterized.
Soil as a multi-phase porous material typical exhibits
different relaxation processes mostly with a distribution of relaxation times
in the frequency range of interest in applications \cite{WagScheu2009a}.
In this context, a broadband numerical 3D FEM study in combination with a
global inversion approach was carried out to analyze
the possibility and limitations for the characterization and parametrization
 of the relaxation behavior.
 To quantify the uncertainty in the estimation of relaxation
parameters from a data-set of full two port scattering parameters in the
frequency range from 1~MHz to 10~GHz known dielectric relaxation spectra of
standard materials as well as soil-related spectra were used. The spectra were
analyzed by means of a broadband generalize dielectric relaxation model (GDR,
see \cite{Wagner2011}).

\section*{\large{2 MODELING AND INVERSION}}
The structure used in the numerical 3D-FEM calculation of the full S-parameter
set was based on a 50~$\Omega$ coaxial transmission line cell introduced in
\cite{Lauer2012} for determining HF-EM properties of undisturbed soil samples.
The cell has an outer diameter of the inner conductor $d_i$=16.9~mm, an inner
diameter of the outer conductor $d_o$=38.8~mm and a length $l$=50~mm. The used
input spectra are choosen according to (i) experimental results of de-ionized
water, ethanol, methanol as well as water with defined electrical conductivity
obtained with open ended coaxial line technique (see \cite{Wagner2013}) and
(ii) different soil types with defined variation in dispersion and absorption
as well as conductive losses (see \cite{Wagner2011, Lauer2012}). To study the
accuracy in determination of relaxation parameters from a data-set of broadband
full two port S-Parameters uni-, bi- and tri-modal dielectric relaxation
spectra are used. In Table \ref{tab:relaxPars} the model parameters of the used
materials are summarized.

\subsection*{2.1 3-D Finite Element Modeling}

The numerical calculation was carried out based on a commercial software
package for 3D-FEM simulation provided by Ansys (Ansoft HFSS). Ansoft HFSS
solves Maxwell's equations using a finite element method, in which the solution
domain is divided into tetrahedral mesh elements. With tangential element basis
functions field values from both nodal values at vertices and on edges are
interpolated. Outer surface cross sections of the coaxial line structure
corresponding to appropriate measurement planes were used as wave ports. A
finite conductivity boundary condition was used at the outer surface in
parallel to the length of the cell. The mesh generation was performed
automatically with l/3 wavelength based adaptive mesh refinement at a solution
frequency of 10~GHz. Broadband complex S-parameters were calculated based on an
interpolating sweep (1~MHz - 10~GHz) with extrapolation to DC.  The computed
generalized S-matrix is normalized to 50~$\Omega$ for comparison of the
numerical calculated with measured S-parameters (see \cite{Wagner2010}).
\begin{table}\label{tab:fit_data}
  \caption{\emph{Relaxation parameters of the assumed input spectra with $\sigma_W$ -  \{0.0004/ 0.010/ 0.04/ 0.25/ 1.15\} S/m,
  $\sigma_S$ - \{0.001/ 0.01/ 0.1/ 1\} S/m, $\Delta\varepsilon_{S}$  - \{3/ 10\}, $\alpha_S$ - \{0.5\}.}}\label{tab:relaxPars}
 \begin{center}
 \begin{tabular}{|l|l|l|l|l|l|l|}
  \hline
  Material & ethanol & methanol & water  & soil \#1  & soil \#2  & soil \#3 \\
   &  &  &  & (sand-analog) & (silt-analog)  & (clay-analog) \\
  \hline
  \hline
  $\varepsilon_\infty$    &    3.9    &  5.65  &  4.5      &  5     &  5        & 5       \\
  $\Delta\varepsilon_{1}$ &    20.7   &  28.0  &  75.2     & 30     & 30        & 30        \\
  $\tau_1$ [ps]&               160.7  &  56.4  &  9.8      &  1    &  1         & 1       \\
  $\alpha_1$ &                 1      &  1   &  1        &    1     & 1         & 1        \\
  $\Delta\varepsilon_{2}$ & -  &    - &    - &   - & $\Delta\varepsilon_{S}$ &  $\Delta\varepsilon_{S}$ \\
  $\tau_2$ [ns]&            -         &    - &    -      &   -      & 10        &  10       \\
  $\alpha_2$ &              -         &    - &    -      &   -      & 1         &  $\alpha_S$    \\
  $\beta_2$ &               -        &    - &   -      &   -    & 0        &  0            \\
  $\Delta\varepsilon_{3}$ & -         &    - &    -      &   -      & -         &  100     \\
  $\tau_3$ [$\mu$s]&        -         &    - &    -      &   -      & -         &  10       \\
  $\alpha_3$ &              -         &    - &    -      &    -     & -         &  $\alpha_S$       \\
  $\beta_3$ &               -         &    - &    -      &    -     & -         &  0         \\
  $\sigma_{DC}$ [S/m]&      0         &  0  & $\sigma_W$ & $\sigma_S$ & $\sigma_S$  &  $\sigma_S$  \\
  \hline
 \end{tabular}\vspace{0.2cm}\\
\end{center}
\end{table}

\subsection*{2.2 Quasi-analytical Inversion}
\begin{figure}[t]
\center
  \includegraphics[scale=0.4]{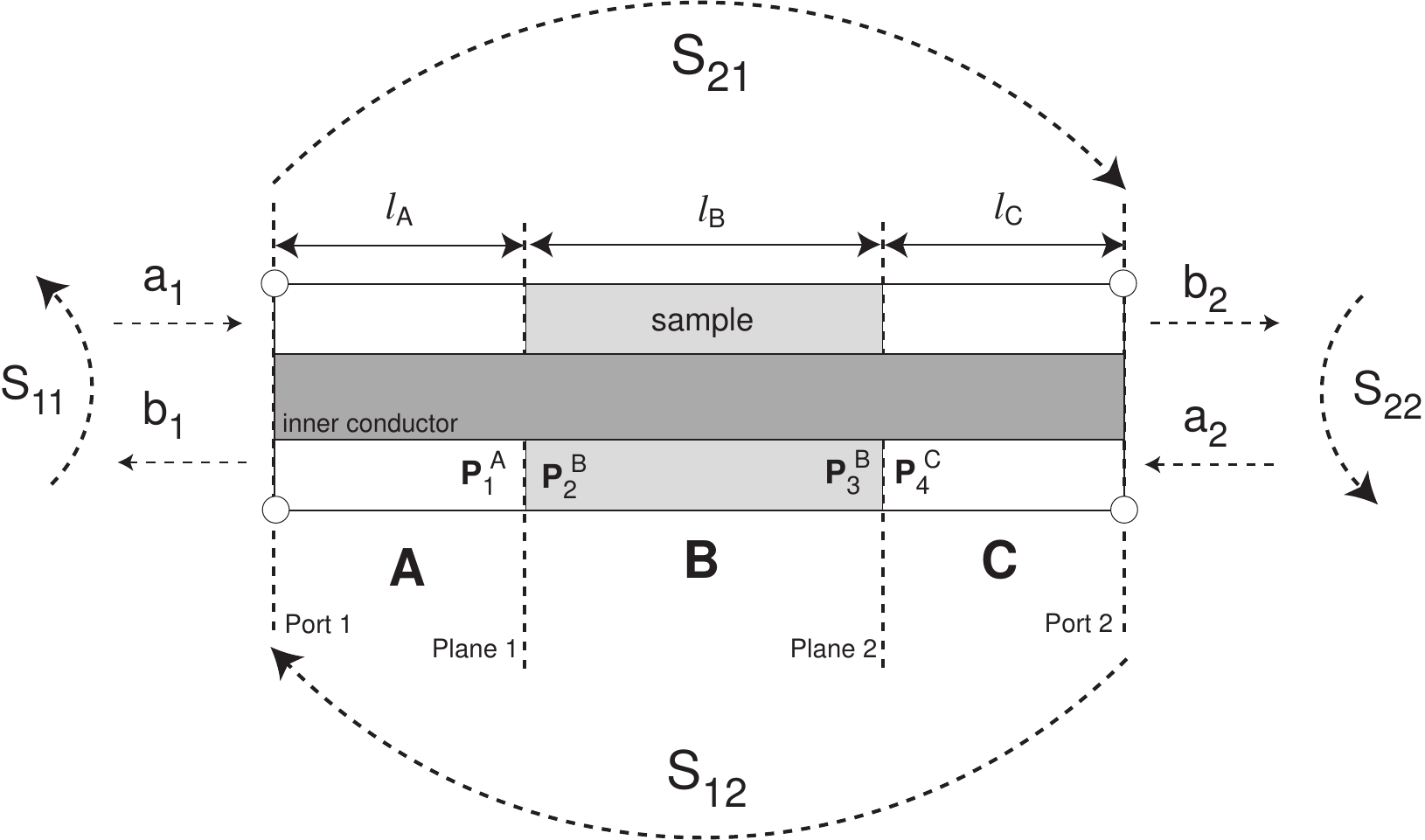}
  \caption{\emph{Schematic illustration of the measurement planes of a coaxial transmission line probe. Port 1 and port 2
    denote measurement planes and Plane 1 and Plane 2 calibration reference plane positions. $S_{ij}$ are the elements of the measured
    complex scattering matrix $\mathbf{S}$ and $a_i$ as well as $b_i$ with $i = 1, 2$ are incident and reflected waves, respectively,
    whereby $i = 1$ and $2$ corresponds
    to the appropriate port connected to the network analyzer. $\mathbf{A}$, $\mathbf{B}$ and $\mathbf{C}$
    are the propagation matrixes of the appropriate sections with length $l_i$ (see text for details).}}\label{fig:Fig0}
\end{figure}

In general, assuming propagation in TEM mode and non
magnetic materials  $\varepsilon_{r,\mbox{eff}}^\star$ of a sample in a transmission line is related to its
complex impedance $Z_S^\star$  or complex propagation factor $\gamma_S^\star$
as follows (see \cite{Nico70, BJ2004, Gorriti2005b, Wagner2010}):
\begin{eqnarray}
   \varepsilon_{r,\mbox{eff}}^\star & = & \left(\frac{Z_0}{Z^\star_S}\right)^2\label{eq:IM},\\
   \varepsilon_{r,\mbox{eff}}^\star & = & \left(\frac{c_0\gamma^\star_S}{j\omega}\right)^2\label{eq:PMM},\\
   \varepsilon_{r,\mbox{eff}}^\star & = & \frac{c_0 Z_0}{j\omega}\left(\frac{\gamma^\star_S}{Z^\star_S}\right) \label{eq:PM}.
\end{eqnarray}
Herein $Z_0$ is the characteristic impedance of the empty transmission line,
$c_0  = \left( {\varepsilon _0 \mu _0 } \right)^{ - 0.5}$ the velocity of light
with $\varepsilon _0$ absolute dielectric permittivity, $\mu_0$ absolute
magnetic permeability of vacuum, $\omega=2\pi f$ angular frequency and
imaginary unit $j=\sqrt{-1}$ . To obtain $Z_S^\star$ or $\gamma_S^\star$ from
measured complex S-parameters $S_{ij}$ several quasi analytical approaches are
available. In the transmission/reflection approach with coaxial transmission
line technique, a sample is inserted into the coaxial line (see Fig.
\ref{fig:Fig0}). The scattering equations corresponding to the experimental
set-up or the implemented numerical model have to be found from an analysis of
the electric field at the sample interfaces.

\paragraph{(i)} In the classical Nicolson-Ross-Weir algorithm (NRW, \cite{Nico70, Weir74})
the scattering parameters are written as follows:
\begin{eqnarray}\label{eq:NRW1}
s^{NR}_{1,j} = \left(S_{ij}e^{(l_a+l_c)}+ S_{jj}e^{2 l_j}\right)e^{\frac{j\omega}{c_0}}, &
s^{NR}_{2,j} = \left(S_{ij}e^{(l_a+l_c)}- S_{jj}e^{2 l_j}\right)e^{\frac{j\omega}{c_0}}, \hspace{0.5cm}
g_{j}=\frac{1-s^{NR}_{1,j}s^{NR}_{2,j}}{s^{NR}_{1,j}-s^{NR}_{2,j}}
\end{eqnarray}
with $i\neq j$ and  $i,j=\{1,2\}$. The reflection coefficient $\Gamma_j$ and the transmission coefficient $T_j$ then are given by:
\begin{equation}\label{eq:NRW2}
\Gamma_j=g_j\pm\sqrt{g_j^2-1},\hspace{0.5cm}
T_j=\frac{s^{NR}_{1,j}-\Gamma_j}{1-s^{NR}_{1,j}\Gamma_j}.
\end{equation}

\paragraph{(ii)}
In the modified NRW algorithm suggested by Baker-Jarvis et al. (see \cite{BJ2004}, hereafter called BJ) the following formulation of the
appropriate scattering parameters is used:
\begin{eqnarray}\label{eq:BJ1}
s^{BJ}_1=(S_{21}S_{12}-S_{11}S_{22})e^{\left[2\frac{j\omega}{c_0}(l-l_b)\right]},
& s^{BJ}_2=\frac{S_{21}+S_{12}}{2}e^{\left[\frac{j\omega}{c_0}(l-l_b)\right]}
\end{eqnarray}
with length $l=l_a+l_b+l_c$ of the empty coaxial line cell between the calibration planes and
length $l_b$ of the coaxial line section filled with the sample. Than $T$ and $\Gamma$ reads:
\begin{equation}\label{eq:BJ2}
T=\frac{s^{BJ}_1+1}{2s^{BJ}_2}\pm\sqrt{\left(\frac{s^{BJ}_1+1}{2s^{BJ}_2}\right)^2-1}, \hspace{0.5cm}
\Gamma=\pm\sqrt{\frac{s^{BJ}_1-T^2}{s^{BJ}_1T^2-1}}.
\end{equation}

The complex propagation factor $\gamma_S^\star$ and complex impedance $Z_S^\star$ are in both cases given as follows:
\begin{equation}\label{eq:NRW4}
\gamma_S^\star=-\frac{\ln(T)}{l_P}, \hspace{0.5cm}
Z_S^\star=Z_0\left(\frac{\Gamma+1}{1-\Gamma}\right).
\end{equation}
The results of equation (\ref{eq:NRW4}) for the propagation factor are
ambiguous due to the logarithm of the complex number $T$ with
$\ln(T)=\ln(|T|)+j\arg(T)$ where the imaginary part is given by $\phi+2\pi n$
with $n= 0, \pm 1, \pm 2, ...$. The correct $n$ as a function of frequency has
to be numerically determined based on the discontinuity of $\partial_f
\textsf{Im}(-\ln T)$. In this study we applied an one dimensional phase
unwrapping procedure provided in matlab. Moreover, the correct roots in
equation (\ref{eq:NRW2}) or (\ref{eq:BJ2}) were chosen such that
$\textsf{Re}(\gamma_S^\star\geq 0)$, $\textsf{Re}(Z_S^\star)\geq 0$ and $|T|
\leq 1$, $|\Gamma|\leq1$ are satisfied.

\paragraph{(iii)} A further approach were suggested by Gorriti et al. \cite{Gorriti2005a} which is based on appropriate propagation matrixes
($\mathbf{A}, \mathbf{B},  \mathbf{C}$ with $\mathbf{\hat{A}}=\mathbf{C},
\mathbf{\hat{B}}=\mathbf{B},  \mathbf{\hat{C}}=\mathbf{A}$, see
\cite{Zhang2007}):
\begin{equation}\label{eq:PM1}
\left[\begin{matrix}1 \\ S_{11} \end{matrix}\right]=\mathbf{A}\mathbf{B}\mathbf{C} \left[\begin{matrix} S_{21} \\ 0 \end{matrix}\right], \hspace{0.5cm}
\left[\begin{matrix}1 \\ S_{22} \end{matrix}\right]=\mathbf{\hat{A}}\mathbf{\hat{B}}\mathbf{\hat{C}} \left[\begin{matrix} S_{12} \\ 0 \end{matrix}\right].
\end{equation}

In the numerical implemented structure section $\mathbf{A}$ and $ \mathbf{C}$
have identical propagation properties and the material in the sample section
$\mathbf{B}$ is homogeneous and isotropic. The determination of the appropriate
complex impedance or propagation factor of the sample from measured scattering
parameters in the PM-approach is here developed in terms of $S_{11}$ and
$S_{21}$. The matrixes are given as follows:
\begin{equation}\label{eq:PM3}
\mathbf{A}=(\mathbf{P}_1^A)^{-1}\left[\prod_{k=2}^{k=p-1}\mathbf{P}_k^C(\mathbf{P}_k^A)^{-1}\right], \hspace{0.5cm}
\mathbf{B}=\mathbf{P}_B^C(\mathbf{P}_B^A)^{-1}, \hspace{0.5cm}
\mathbf{C}=\left[\prod_{k=p+1}^{k=N-1}(\mathbf{P}_k^A)^{-1}\mathbf{P}_k^C\right]\mathbf{P}_N^C
\end{equation}
with interface between section $k$ and $k+1$ as well as
\begin{equation}\label{eq:PM6}
\mathbf{P}_k^A=\left[\begin{matrix}\exp(-\gamma_k^\star l_k) & 1 \\
\frac{1}{Z_k}\exp(-\gamma_k^\star l_k) & -\frac{1}{Z_k^\star}
\end{matrix}\right], \hspace{0.5cm}
\mathbf{P}_{k}^C = \left[\begin{matrix}1 & \exp(-\gamma_{k}^\star l_{k}) \\
\frac{1}{Z_{k}^\star} & -\frac{1}{Z_{k}^\star}\exp(-\gamma_{k}^\star l_{k})
\end{matrix}\right]
\end{equation}
Herein $\gamma_k^\star$ are the appropriate propagation factor and $Z_k^\star$
characteristic impedance of the $k$-th section with sample section \textbf{B}
labeled with $p$. The dielectric relaxation spectrum of the material under
study is contained within matrix \textbf{B}, which is given with equation
(\ref{eq:PM3}) and (\ref{eq:PM6}) by
\begin{equation}\label{eq:TMLI2}
\mathbf{B}=\frac{1}{2}\left[\begin{matrix}
e^{\gamma_p^\star l_p}+e^{-\gamma_p^\star l_p} & Z_p^\star\left[e^{\gamma_p^\star l_p}-e^{-\gamma_p^\star l_p}\right]  \\
\frac{1}{Z_p^\star}\left[e^{\gamma_p^\star l_p}-e^{-\gamma_p^\star l_p}\right]
& e^{\gamma_p^\star l_p}+e^{-\gamma_p^\star l_p}
\end{matrix}\right].
\end{equation}
Equation (\ref{eq:PM1}) in terms of $S_{11}$ and $S_{21}$ can be rewritten
\begin{equation}\label{eq:TMLI3}
\left[\begin{matrix} a \\ b \end{matrix}\right]=
\mathbf{B}\left[\begin{matrix} c \\ d \end{matrix}\right], \hspace{0.5cm}
\left[\begin{matrix} a \\ b \end{matrix}\right]=
\mathbf{A}^{-1}\left[\begin{matrix} 1 \\ S_{11} \end{matrix}\right], \hspace{0.5cm}
\left[\begin{matrix} c \\ d \end{matrix}\right]= \mathbf{C}\left[\begin{matrix}
S_{21} \\ 0 \end{matrix}\right].
\end{equation}
Substituting $\mathbf{B}$ according to equation (\ref{eq:TMLI2}) into equation
(\ref{eq:TMLI3}) and eliminating the exponential terms, the impedance
$Z_S^\star$ of the sample than becomes
\begin{equation}\label{eq:TMLI6}
Z_S^{\star}=\pm\sqrt{\frac{a^2-c^2}{b^2-d^2}}.
\end{equation}
If the impedance terms are eliminated the propagation factor is given as
\begin{equation}\label{eq:TMLI8}
\gamma_S^\star=\pm a\cosh\left(\frac{ab+cd}{ad+bc}\right)l^{-1}.
\end{equation}
The correct sign were chosen such that $\textsf{Re}(Z_S^\star)\geq 0$ and $\textsf{Re}(\gamma_S^\star\geq 0)$.

\begin{figure}[t]
  \includegraphics[scale=0.55]{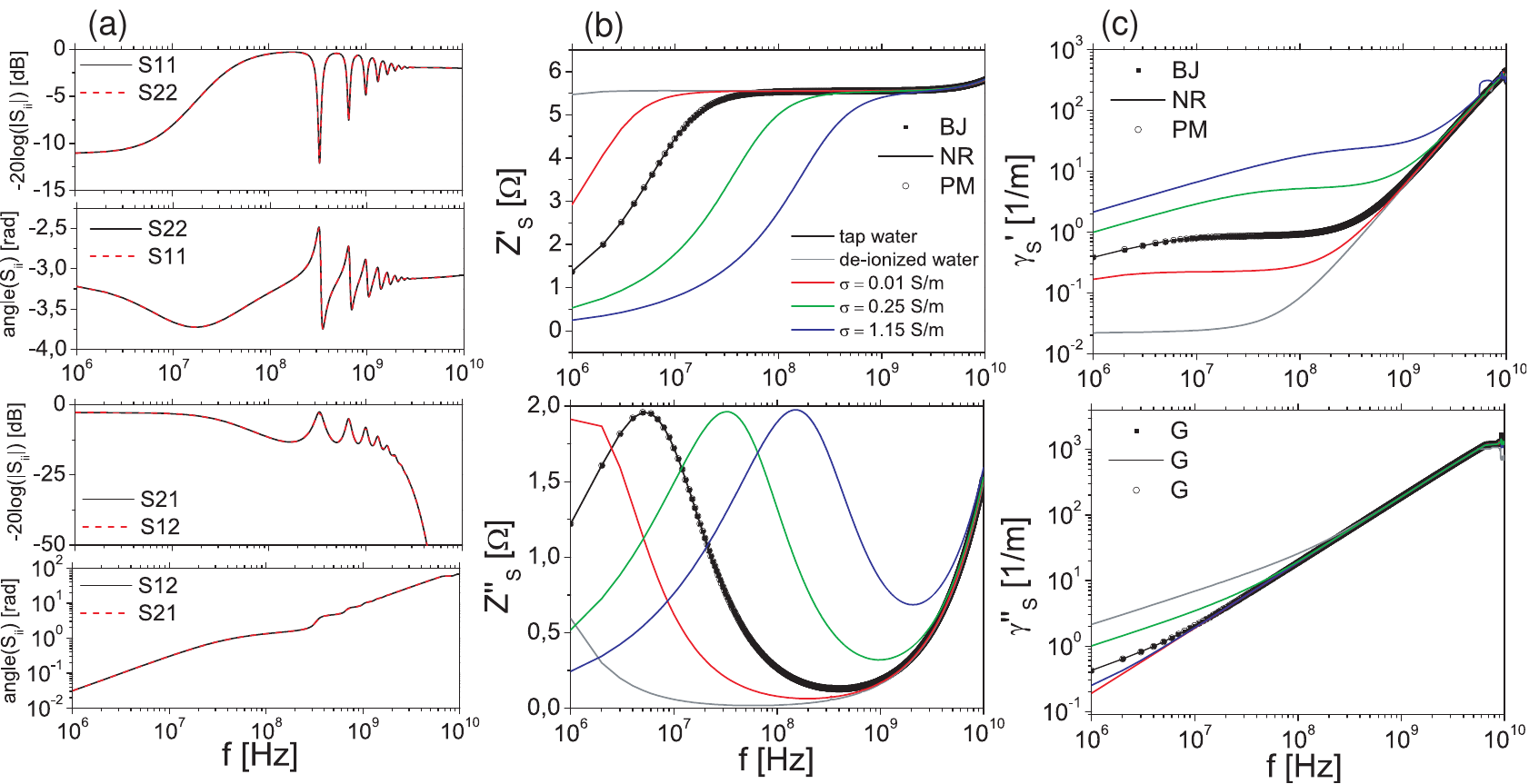}
  \caption{\emph{Numerical results for the water filled coaxial transmission line (a) reflection $S_{ii}$ and transmission factor $S_{ij}$
  with absolute value and phase, (b) complex impedance $Z^\star_S$ with real $Z'_S$ ans imaginary part $Z''_S$ as well as
  (c) complex propagation factor $\gamma^\star_S$ obtained with the three different approaches.}}\label{fig:Fig2}
\end{figure}

Hence, all three algorithms can be formulated in a way to determine explicitly $Z_S^\star$ and $\gamma_S^\star$.
For the calculation of the permittivity spectra
the appropriate relationship according to equation (\ref{eq:IM}), (\ref{eq:PMM}) or (\ref{eq:PM}) can be used (see \cite{Wagner2011a}).
In Fig. \ref{fig:Fig2} the results in case of tap water
are shown.
The numerical calculated complex S-parameters with the 3D-FEM approach subsequently were used to compute $\varepsilon_{r,\mbox{eff}}^\star$
in the frequency range between 1~MHz to 10~GHz with the introduced quasi-analytical methods:
classical NRW, BJ and the propagation matrix algorithm (PM).

\subsection*{2.3 Inverse Modeling}

Soil as a multi-phase porous material typical exhibits several relaxation
processes in the frequency range between 1~MHz and 10~GHz of interest
in applications \cite{Robinson2008, Robi03, Wagner2011}. Dielectric loss spectra of
saturated and unsaturated porous materials are the result of broadly distributed
relaxation processes (\cite{Hoek74, Holl98, Ishi00, Kell05, Wagn07, Wagn07a, Wagner2011, Wagner2013}).
Based on the theory of fractional time evolutions Hilfer \cite{Hilf02} derived
relaxation functions for the complex frequency dependent
dielectric permittivity of amorphous and glassy materials which are used to develop an
generalized broadband relaxation model (GDR, see \cite{Wagn07a, Wagner2011}):
\begin{equation}\label{eq:GDR}
\varepsilon_{\mbox{r,eff}}^\star-\varepsilon_\infty= \sum\limits_{k = 1}^N
{\frac{{\Delta\varepsilon _k }}{{\left( {j\omega \tau _k} \right)^{\alpha _k} +
\left( {j\omega \tau _k} \right)^{\beta _k} }}} - j\frac{{\sigma
_{DC}}}{{\omega \varepsilon _0 }}
\end{equation}
with high frequency limit of permittivity $\varepsilon_\infty$, relaxation
strength $\Delta\varepsilon_k$, relaxation time $\tau_k$ as well as stretching
exponents $0\leq\alpha_k, \beta_k\leq 1$ of the $k$-th process and apparent
direct current electrical conductivity $\sigma_{DC}$.

The GDR-model was fitted to the dataset using a~shuffled complex
evolution metropolis algorithm (SCEM-UA, \cite{Vrugt2003}). This
algorithm is an adaptive evolutionary Monte Carlo Markov chain method and
combines the strengths of the Metropolis algorithm, controlled random search, competitive
evolution, and complex shuffling to obtain an efficient estimate of the most optimal parameter set, and its
underlying posterior distribution, within a~single optimization run \cite{Heimo04}.

The algorithm is based on a~Bayesian inference scheme. The needed prior
information are a~lower and upper bound for each of the relaxation parameters
${\xi}$. Assuming this non informative prior the posterior density ${p(\xi
\vert y,\gamma)}$ for ${\xi}$ conditioned with the measurement $y$ is given by
\cite{Vrugt2003}:
\begin{equation}
\label{eq19} {p(\xi \vert y,\gamma)\propto }\left[ \sum\limits_{{k=1}}^{N}
\frac{{y}_{{k}}{-}\hat{y}_{{k}}}{{\delta }} \right]^{{-N(1+\gamma)/2}}.
\end{equation}
\begin{figure}[h]
\center
  \includegraphics[scale=0.73]{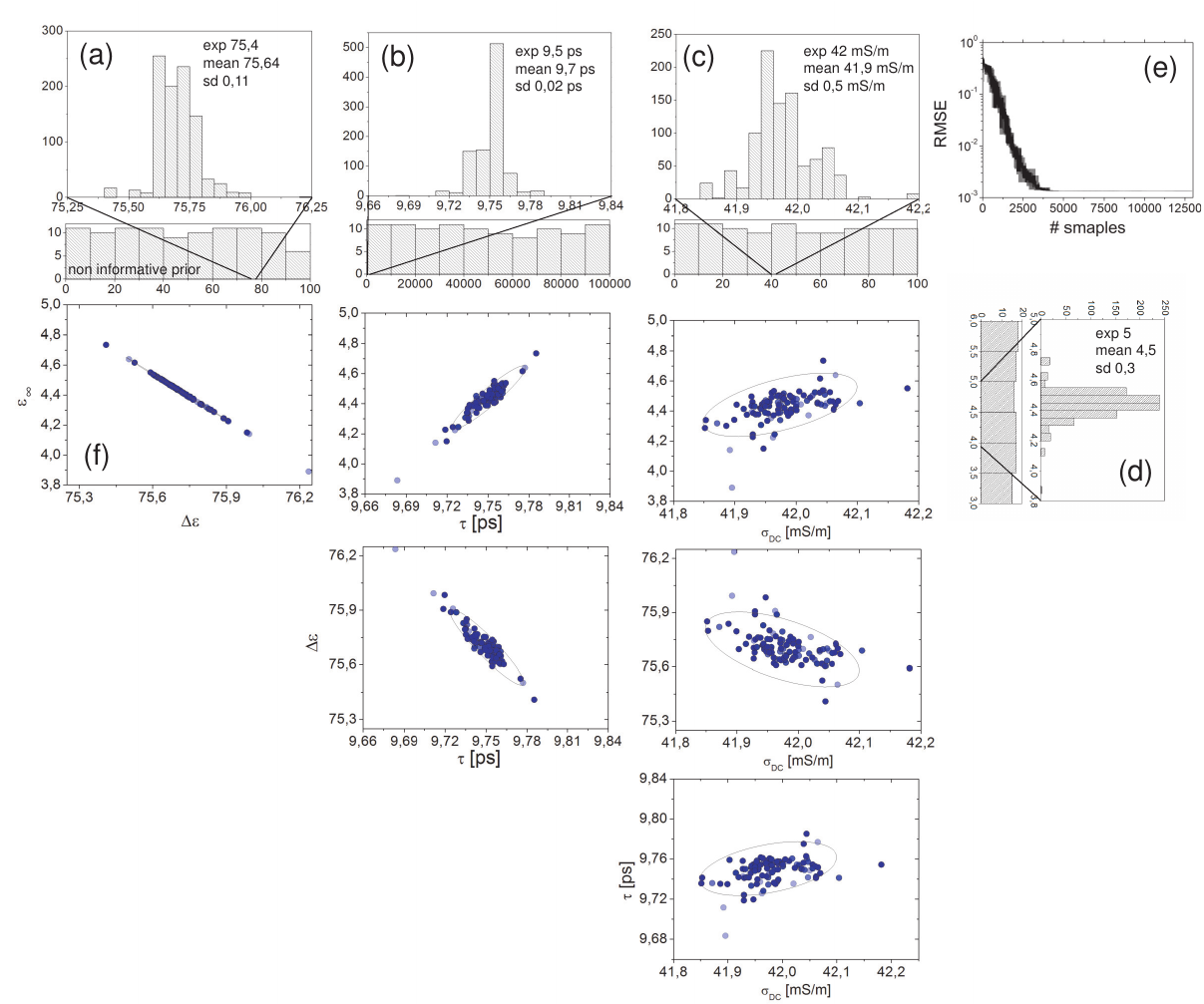}
  \caption{\emph{Results of the SCEM-UA optimization with a one element GDR (Debye-type) in case
           of a coaxial line cell filled with tap-water.
           (a)-(d) Marginal posterior probability distributions
           of the high frequency limit of permittivity $\varepsilon_\infty$, relaxation strengt $\Delta\varepsilon$, relaxtion time $\tau$ and
           direct current conductivity $\sigma_{DC}$ obtained by means of $1000$ samples generated after convergence to a
           posterior distribution (see RMSE (e)).
           (f) Bivariate scatter plots of the behavioral (posterior) samples.}}\label{fig:Fig3}
\end{figure}
Herein ${y}_{{k}}$ is the $k$-th of $m$ measurements at each frequency and
$\hat{y}_{{k}}$ is the corresponding model prediction. ${ \delta}$ represents
the error of the numerical calculated S-parameters as a~standard deviation. The
parameter ${\gamma}$ specifies the error model of the residuals. In the
implemented numerical model $\delta$-values are less than $-50$~dB. The
residuals are assumed normally distributed when ${\gamma =0}$, double
exponential when ${\gamma =1}$, and tend to a~uniform distribution as ${\gamma
\to -1}$. In this study we use $\gamma=0$.
\begin{figure}[t]
\center
  \includegraphics[scale=0.47]{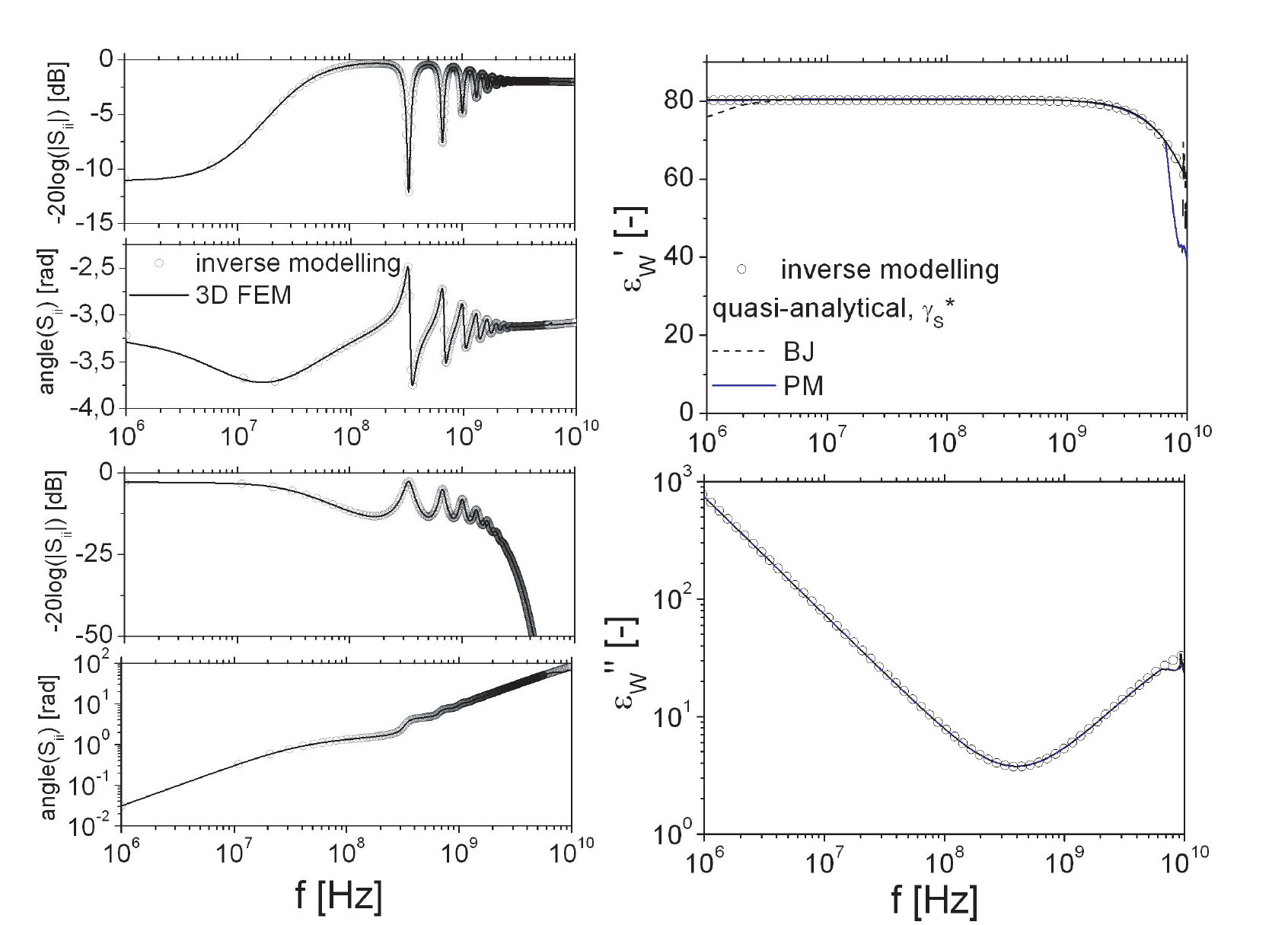}
  \caption{\emph{(left) Numerical determined full S-parameter set $S_{ij}$ in comparison with the result of the
                  inverse modeling and (right) appropriate dielectric spectra obtained with the quasi-analytical
                  propagation factor approach
                  as well as inverse modeling technique.}}\label{fig:Fig4}
\end{figure}
In Fig. \ref{fig:Fig3} optimization results are represented in case of a coaxial transmission line filled with tap-water.
The corresponding spectra obtained with the quasi-analytical approach as well as inverse modeling technique are
represented in Figure \ref{fig:Fig4}.
The optimization results are in close agreement with the expected parameters especially in case of the simple Debye-model.
The appropriate bivariate scatter plots further indicate the expected correlation between $\varepsilon_\infty$ and $\Delta\varepsilon$ due
to the limited accessible frequency range (see Figure \ref{fig:Fig4}).
\begin{figure}[ht]
\center
  \includegraphics[scale=0.26]{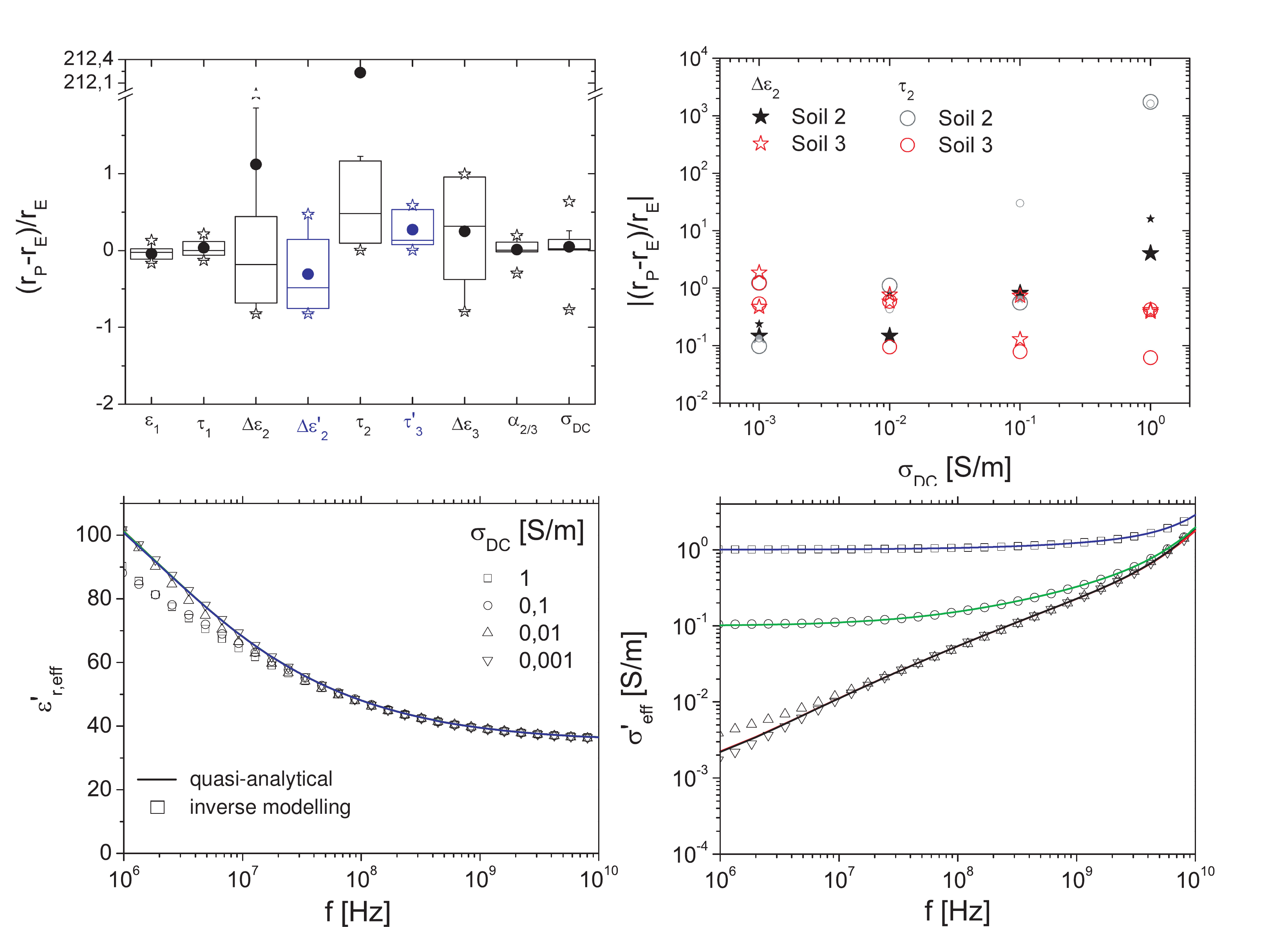}
  \caption{\emph{(top/left) Box and whiskers plots of the relaxation parameters obtained with inverse modeling technique ($r_P$)
            related to the expected parameters ($r_E$). Parameters $\Delta\varepsilon_2'$ and $\tau_2'$ obtained considering prior information
            due to a high conductivity contribution (see text).
            (top/right) relative error of the obtained relaxation parameters for process 2 as a function of
            given direct current conductivity contribution $\sigma_{DC}$. Small symbols were used for $\Delta\varepsilon_{S}=3$.
            (bottom) Determined dielectric spectra based on quasi-analytical as well as inverse modeling technique
            in case of soil \#3 for $\Delta\varepsilon_S=10$, $\alpha_S=0.5$ and varying $\sigma_{DC}$.}\\}\label{fig:Fig5}
\end{figure}

\section*{\large{3 RESULTS AND DISCUSSION}}
In Figure \ref{fig:Fig5} the error statistics of the different relaxation
parameters related to the expected parameters of Table \ref{tab:relaxPars} are
represented for all investigated materials. The relative error of the
relaxation time of process 3 is not included in the plot due to a high
uncertainty related to the restricted frequency range. The error statistics
clearly indicate an reasonable accuracy in the determination of the relaxation
parameters of process 1 as well as the direct conductivity in agreement with
experimental results \cite{Wagner2011, Wagner2012, Wagner2013}. This in
addition shows the results of the obtained spectra in comparison with the
quasi-analytical approaches. However, a serious limitation is observed in case
of low $\sigma_{DC}$ values $<0.01$~S/m leading to an increase in the
uncertainty in the parameter estimation process (Figure \ref{fig:Fig5},
bottom). The uncertainty in the prediction of relaxation parameters
$\Delta\varepsilon_2$ and $\tau_2$ is strongly affected by the conductivity
contribution (Figure \ref{fig:Fig5}, top/right). This suggests to separate
conductivity effects prior to the analysis of the spectra especially in case of
high conductivities $>1$~S/m leading to a substantial decrease in the
uncertainty. Moreover, the magnitude of the process in relationship to further
overlying processes strongly affects the uncertainty of the parameter
estimation process. In contrast, both relaxation time distribution parameters
were estimated with reasonable accuracy.

\section*{\large{4 CONCLUSION}}

The results show: (i) regardless of the underlying relaxation processes, the
dielectric spectra can be determined in close agreement with the assumed input
spectra and (ii) the relaxation behavior can be parameterized with reasonable
accuracy if the relaxation frequency $f_{r,i}$ of the single processes is in
the frequency range of the measurement. If $f_{r,i}$ is below or above the
measurement range, than the accuracy of the estimated relaxation parameters
strongly decreases. In context of spectra related to porous media it was found
that the relaxation behavior of the free water contribution (primary
$\alpha$-process) as well as the apparent conductivity contribution
$\sigma_{DC}$ can be characterized in close agreement with experimental
results. The low frequency $\beta$ process accounts for the low frequency
dispersion as well as electrode polarization effects which is the most
challenging process. The secondary $\alpha'$ process can be characterized, but
the relaxation characteristics is strongly affected by the overlapping
$\beta$-process as well as the magnitude of the direct current conductivity
contribution. To quantitatively relate the relaxation parameters to soil
physico-chemical quantities (i) the measurement range have to be extended to
lower frequencies, (ii) an informative prior have to be included in the
optimization process, and (iii) the complexity of the relaxation model has to
be constraint by means of a combination of relaxation models and mixture
equations. In next steps, the results determined in parameterizing the
 the S-parameters will be compared with results obtained by
classical parametrization of the dielectric spectra with quasi-analytical
approaches based on the Geophysical Inversion and Modelling Library (GIMLi, see
\cite{Loewer2013}).



\small
\bibliographystyle{ieeetr}
\bibliography{Literatur1}

\end{document}